\documentclass[superscriptaddress,prd,showpacs,aps]{revtex4}
\usepackage{amsmath}
\usepackage{graphicx}
\usepackage{verbatim}

\begin{document}

\newcommand{\HC}{{\rm H.c}}
\newcommand{\TeV}{{\rm TeV}}
\newcommand{\GeV}{{\rm GeV}}
\newcommand{\CKM}{{\rm CKM}}
\def\met{\displaystyle{\not}E_T}

\title{Identify Charged Higgs Boson in $W^\pm H^\mp$ associated production at LHC}

\author{Shou-Shan Bao}\email{ssbao@sdu.edu.cn}\affiliation{School of Physics, Shandong University, Jinan Shandong 250100, P.R.China}
\author{Xue Gong}\email{gongxue@mail.sdu.edu.cn}\affiliation{School of Physics, Shandong University, Jinan Shandong 250100, P.R.China}
\author{Hong-Lei Li}\email{lihl@mail.sdu.edu.cn}\affiliation{School of Physics, Shandong University, Jinan Shandong 250100, P.R.China}
\author{Shi-Yuan Li}\email{lishy@sdu.edu.cn}\affiliation{School of Physics, Shandong University, Jinan Shandong 250100, P.R.China}
\author{Zong-Guo Si}\email{zgsi@sdu.edu.cn}\affiliation{School of Physics, Shandong University, Jinan Shandong 250100, P.R.China}\affiliation{Center for High-Energy Physics, Peking University, Beijing 100871, P.R.China}

\begin{abstract}
We investigate the possibility to discover the charged
Higgs via $pp\to W^{\pm}H^{\mp}\to l+\met+b\bar{b}jj$ process at
LHC, which suffers from large QCD backgrounds.
We optimize the kinematic cuts to suppress the backgrounds, so
that the reconstruction of the charged Higgs through hadronic decay
is possible. The angular distribution of the b-jet from $H^{\pm}$
decay is investigated as a way to identify the charged scalar from
vector bosons. 
\end{abstract}

\pacs{12.60.Fr; 14.80.Fd; 14.65.Ha}
\maketitle

\section{Introduction}

Understanding electroweak symmetry breaking is one of the driving
forces behind the undergoing experiments at the CERN large hadron
collider (LHC). In the Standard Model (SM), the fermions and gauge bosons
get masses through Higgs mechanism with one weak-isospin doublet
Higgs field. Although SM is extremely successful in phenomenology, there
are still remaining problems not well understood.
Extensions of SM have been considered widely. Two Higgs Doublet
Model (2HDM)\cite{TDLEE,SW,Liu:1987ng,YLW1,Glashow:1976nt,HW} is one of the natural extensions.
In this kind of models, the charged Higgs boson ($H^{\pm}$) is of
special interest, since its discovery is an unambiguous evidence for an extended Higgs sector.
Therefore, the hunt for charged Higgs bosons plays an important role
in the search for new physics at LHC.

 Currently most of the limits or constraints to the charged Higgs
mass are model-dependent. The best model-independent direct limit
from the LEP experiments is $m_{H^{\pm}}>78.6$ GeV at 95\% C.L.\cite{lep:2001ch},
assuming only the decay channels $H^{+}\to c\bar{s}$ and $H^{+}\to\tau\nu_{\tau}$.
As the charged Higgs will contribute to flavor changing neutral currents at one loop level, the indirect constraints can be extracted
from B-meson decays. In Type II 2HDM, the constraint is $M_{H^{\pm}}\gtrsim350$
GeV for $\tan\beta$ larger than 1, and even stronger for smaller
$\tan\beta$ \cite{typeII}. However, since the phases of the Yukawa
couplings in Type III or general 2HDM are free parameters, $m_{H^{\pm}}$
can be as low as 100 GeV\cite{BCK}.

At hadron colliders, the charged Higgs phenomenology has been studied
widely. The main production modes are $gb\rightarrow H^{-}t$ for $m_{H^{\pm}}>m_{t}+m_{b}$
and $gg\rightarrow H^{-}t\bar{b}$ for $m_{H^{\pm}}\lesssim m_{t}-m_{b}$
\cite{Barnett,Bawa,Borzumati,Miller,Alwall:2004xw,Beccaria:2009my}.
The preferred decay modes are then $H^{-}\rightarrow b\bar{t}$ and
$H^{-}\rightarrow\tau\bar{\nu}_{\tau}$, respectively. Another interesting
channel is the $H^{\pm}$ production in association with a $W$ boson,
whose leptonic decays can serve as an important
trigger for the $W^{\pm}H^{\mp}$ search. This channel can also cover
the region $m_{H^{\pm}}\sim m_{t}$. The dominant channels for $W^{\pm}H^{\mp}$
production are $b\bar{b}\rightarrow W^{\pm}H^{\mp}$ at tree level
and $gg\rightarrow W^{\pm}H^{\mp}$ at one-loop level. $W^{\pm}H^{\mp}$
production at hadron colliders in Type II 2HDM and other 2HDMs has
been studied in \cite{Kniehl,Dicus,Brein,Asakawa,Eriksson,Bao:2010sz}.
It is found that, due to the large negative interference term between the
triangle- and box-type quark-loop Feynman diagrams, the gluon-fusion cross section in MSSM is quite small.
The NLO-QCD and SUSY-QCD corrections to $b\bar{b}$ annihilation amplitude
in MSSM are all about 10\%\cite{Zhao:2005mu,Gao:2007wz}. As has studied
in \cite{Kniehl,Moretti:1998xq}, the hadronic decay channel of $H^{\pm}$
suffers from large QCD backgrounds, which overwhelm the charged Higgs
signal over the heavy mass range that can be probed at LHC. 
In this paper, we optimize the kinematic cuts of the final state. The signal-background
ratio is improved so that the resonance reconstruction is possible. 
 Once the resonance of $b\bar t~(t\bar b)$ is detected at LHC, its spin 
 should be determined. We
investigate the angular distribution of the b-jet with respect to the
beam direction. It is found that such distribution is useful to identify
the charged scalar from vector bosons.

This paper is organized as follows. In section II, the corresponding
theoretical framework is briefly introduced. Section III is devoted to the numerical analysis of $W^{\pm}H^{\mp}$ production and the related SM backgrounds.  In section
IV, the angular distributions of b-jet are investigated to identify the scalar from vector bosons. Finally, a short summary is given.

\section{Theoretical framework for Charged Higgs}
\begin{figure}
\begin{centering}
\begin{tabular}{cc}
\includegraphics[clip,width=4cm,height=10cm,keepaspectratio]{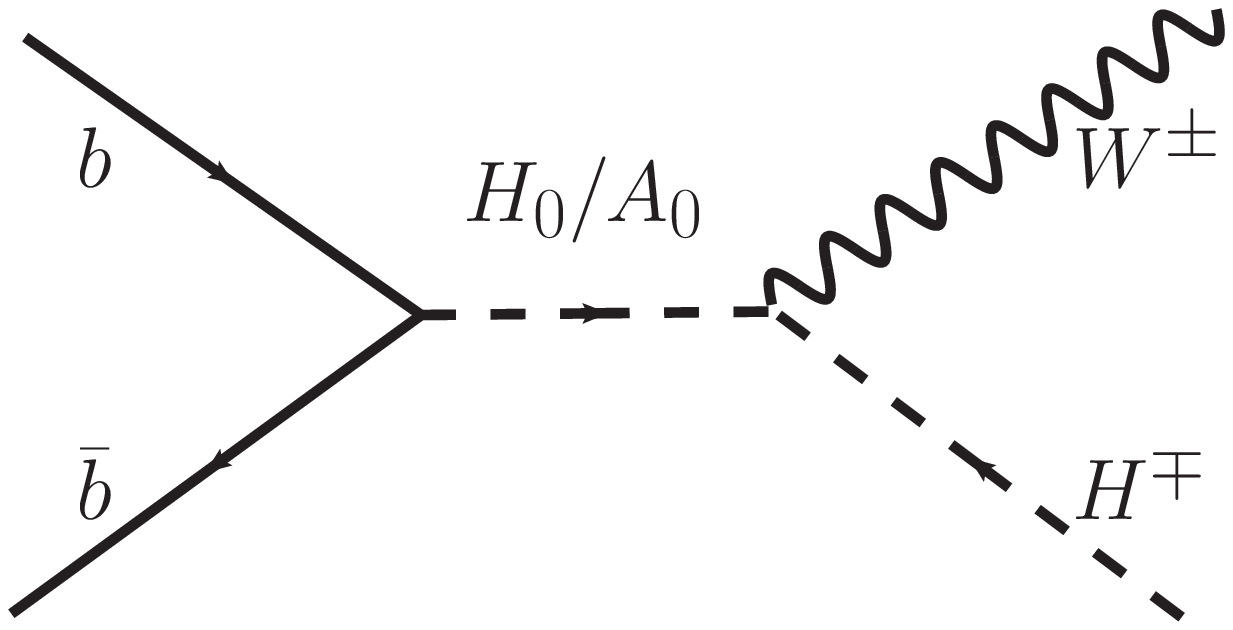} & \includegraphics[clip,width=3.5cm,height=3cm,keepaspectratio]{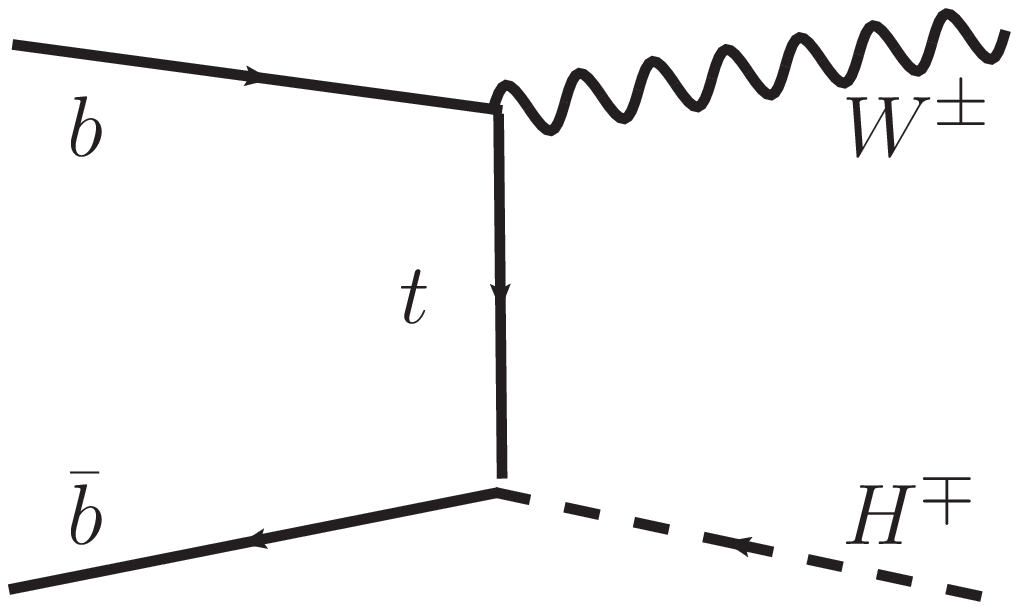}\tabularnewline
(a) & (b)\tabularnewline
\end{tabular}\caption{Feynman diagrams for $W^{\pm}H^{\mp}$ production at partonic level.}
\label{fig:Fdiagram} 
\par\end{centering}
\centering{}
\end{figure}

The scalar sector of SM is not yet confirmed by experiments and
it is possible to extend the Higgs structure to two Higgs doublets. For example,
it has been shown that if one Higgs doublet is needed for the mass
generation, an extra Higgs doublet is necessary for the Spontaneous
$CP$ violation\cite{TDLEE}. In 2HDMs, after the
spontaneous symmetry breaking, there remain five physical Higgs scalars, i.e.,
two neutral $CP$-even bosons $h_{0}$ and $H_{0}$, one neutral $CP$-odd
boson $A_{0}$, and two charged bosons $H^{\pm}$. In this work we aim to study the charged
Higgs phenomenology and choose the Type II Yukawa couplings as the
working model, 
\begin{eqnarray}
-\mathcal{L} & = & -\cot\beta\frac{m^{u}}{v}\bar{u}_{L}(H+iA)u_{R}-\cot\beta\frac{m^{u}}{v}\bar{u}_{R}(H-iA)u_{L}\nonumber \\
 &  & +\tan\beta\frac{M^{d}}{v}\bar{d}_{L}(H-iA)d_{R}+\tan\beta\frac{M^{d}}{v}\bar{d}_{R}(H+iA)d_{L}\nonumber \\
 &  & -\sqrt{2}\cot\beta\frac{M^{u}}{v}V_{ud}^{\dag}\bar{d}_{L}H^{-}u_{R}-\sqrt{2}\tan\beta\frac{M^{d}}{v}V_{ud}^{\dag}\bar{d}_{R}H^{-}u_{L}\nonumber \\
 &  & -\sqrt{2}\cot\beta\frac{M^{u}}{v}V_{ud}\bar{u}_{R}H^{+}d_{L}-\sqrt{2}\tan\beta\frac{M^{d}}{v}V_{ud}\bar{u}_{L}H^{+}d_{R}. 
\label{lagrangian}
\end{eqnarray}
 The Yukawa couplings are related to the fermion masses, which indicates
that the dominant production processes for the charged Higgs
associated with a $W$ boson at hadron collider are $b\bar{b}$ annihilation
at tree-level and gluon fusion at one loop level. The $b\bar{b}$
annihilation is overwhelming for $\tan\beta>1$\cite{Kniehl}, while
large $\tan\beta$ is favored by $B$ meson rare decays\cite{Hewett:1992is,Barger:1992dy,Bertolini:1990if}.
In this work we investigate the properties of charged Higgs with large
$\tan\beta$ in $b\bar{b}$ annihilation process. The Feynman diagrams
are shown in Fig.~\ref{fig:Fdiagram}. The corresponding invariant amplitude
square averaged over the spin and color of initial partons is given
by 
\begin{eqnarray}
|{\cal M}|^{2} & = & \left.\frac{G_{F}^{2}{\hat{s}}}{6}\right\{ 2m_{b}^{2}\tan^{2}\beta[({\hat{s}}-m_{W}^{2}-m_{H}^{2})^{2}-4(m_{W}m_{H})^{2}](|G_{H_{0}}|^{2}+|G_{A_0}|^{2})\nonumber \\
 & + & [m_{b}^{2}\tan^{2}\beta({\hat{s}}{\hat{t}}^{2}+2{\hat{t}}{\hat{u}}m_{W}^{2}-2m_{W}^{4}m_{H}^{2})+m_{t}^{4}\cot^{2}\beta({\hat{t}}{\hat{u}}+2{\hat{s}}m_{W}^{2}-m_{W}^{2}m_{H}^{2})]\frac{4}{{\hat{s}}({\hat{t}}-m_{t}^{2})^{2}}\nonumber \\
 & + & \left.m_{b}^{2}\tan^{2}\beta({\hat{t}}^{2}+{\hat{t}}{\hat{u}}-2m_{W}^{2}m_{H}^{2})\frac{1}{{\hat{t}}-m_{t}^{2}}Re(G_{H_{0}}+G_{A_0})\right\},
\label{msquare}
\end{eqnarray}
 where $\hat{s}=(p_b+p_{\bar b})^2,~\hat{t}=(p_b-p_W)^2,~\hat{u}=(p_b-p_H)^2$ are the Mandelstam variables and
$G_{F}$ is the Fermi constant. The Higgs propagator functions are

\begin{eqnarray}
G_{H_{0},A_0}=\frac{1}{\hat{s}-m_{H_{0},A_0}^{2}+i\Gamma_{H_{0},A_0}m_{H_{0},A_0}},
\end{eqnarray}
where $\Gamma_{H_{0},A_0}$ are the higgs widths which are obtained with HDECAY\cite{Djouadi:1997yw} package.

\section{$W^{\pm}H^{\mp}$ Production and Corresponding backgrounds}\label{signal} 
\begin{figure}
\begin{centering}
\begin{tabular}{cc}
\includegraphics[clip,width=6cm,height=6cm,keepaspectratio]{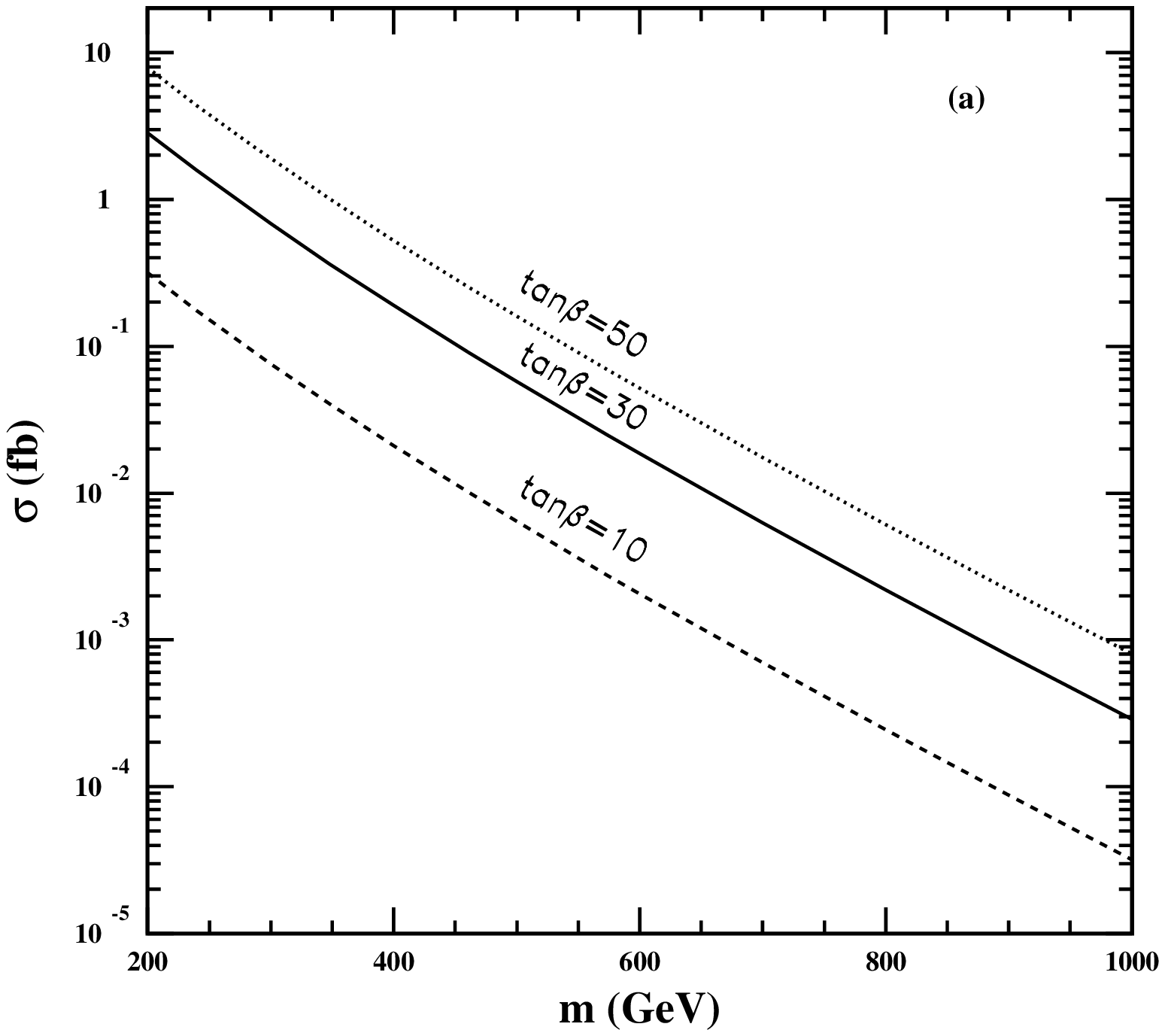} & \includegraphics[clip,width=6cm,height=6cm,keepaspectratio]{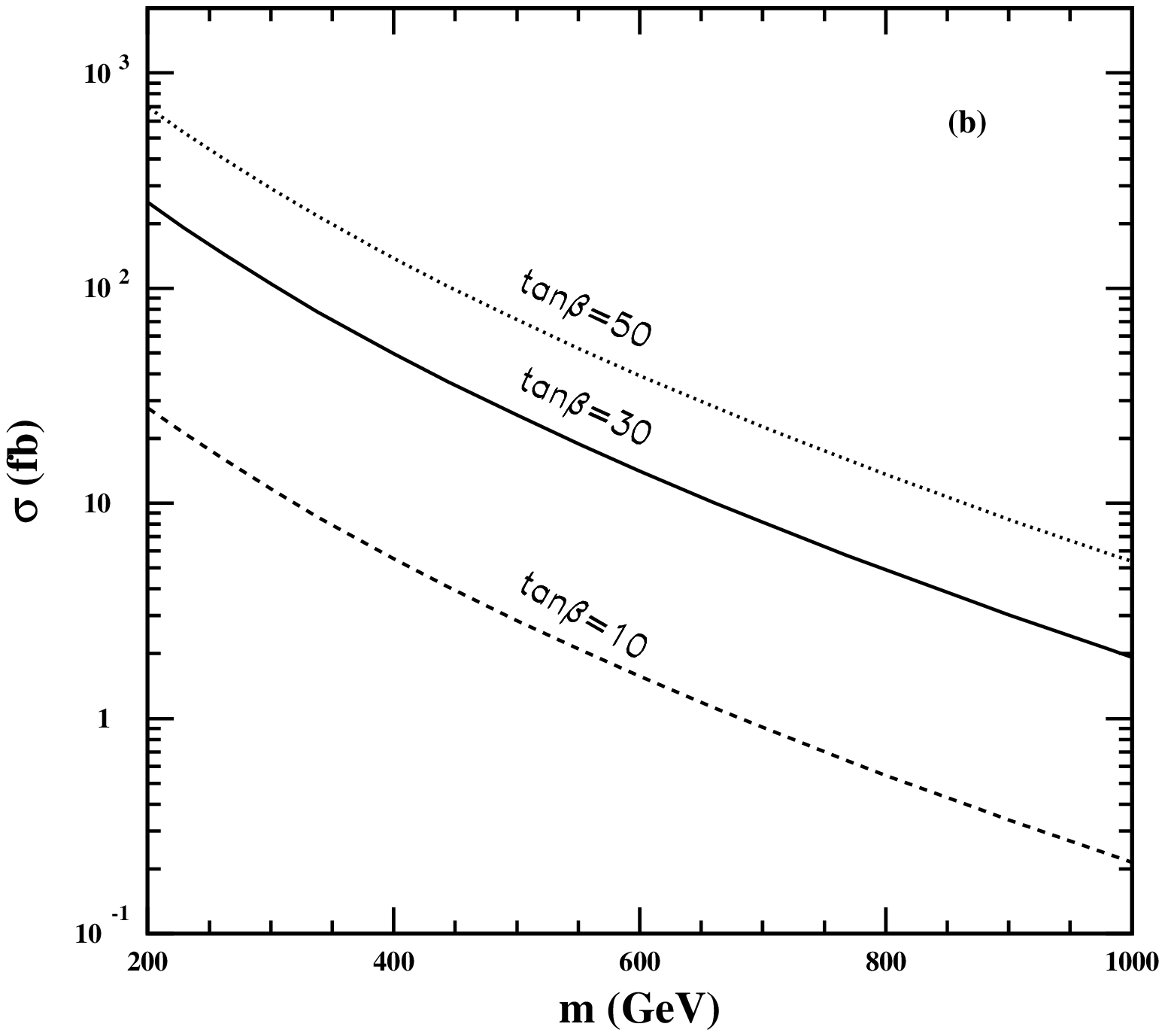}\tabularnewline
\end{tabular}\caption{The total cross section as a function of $m_{H^{\pm}}$ for $pp\to W^{\pm}H^{\mp}$
process at (a) $\sqrt{s}=7$ TeV and (b) $\sqrt{s}=14$ TeV.}
\label{fig:cs}
\par\end{centering}
\centering{} 
\end{figure}

The total cross section for the $pp\to W^{\pm}H^{\mp}$ process can
be written as follows 
\begin{equation}
\sigma=\int f_{b}(x_{1})f_{\bar{b}}(x_{2})\hat{\sigma}(x_{1}x_{2}s)dx_{1}dx_{2},
\end{equation}
where $\sqrt{s}$ is the proton-proton center of mass energy, $\hat{\sigma}$ is the partonic level cross section of $b\bar b\to W^{\pm}H^{\mp}$, and $f_{q}(x_{i})$
is the parton distribution function (PDF). In our numerical calculations,
we employ CTEQ6L1\cite{Pumplin:2002vw} for PDF, and set $V_{tb}=1$,
$M_{W}=80.4$ GeV, and $m_{t}=172.9$ GeV. In Fig.~\ref{fig:cs},
the total cross sections are shown as a function of charged Higgs mass
for $\tan\beta=10$, 30, and 50 at LHC. The cross section increases with $\tan\beta$. 
Supposing the luminosity to be $10~fb^{-1}$
at $\sqrt{s}=7$ TeV, one can notice that it is difficult for the charged
Higgs associated with a $W$ boson to be detected when
its mass is above 600 GeV even for $\tan\beta=50$. It is
easier for the charged Higgs boson to be observed at $\sqrt{s}=14$ TeV. Therefore
in this work we focus on investigating the charged Higgs associated with a $W$ boson in the following processes
\begin{eqnarray}
&pp&\to W^{-}H^{+}\to W^{-}t\bar{b}\to l^{-}\nu b\bar{b}jj, \nonumber \\
&pp&\to W^{+}H^{-}\to W^{+}\bar{t}b\to l^{+}\nu b\bar{b}jj
\label{pro:decay}
\end{eqnarray}
at $\sqrt{s}=14$ TeV with $\tan\beta=30$. 
\begin{figure}
\begin{centering}
\begin{tabular}{ccc}
\includegraphics[clip,width=6cm,height=6cm,keepaspectratio]{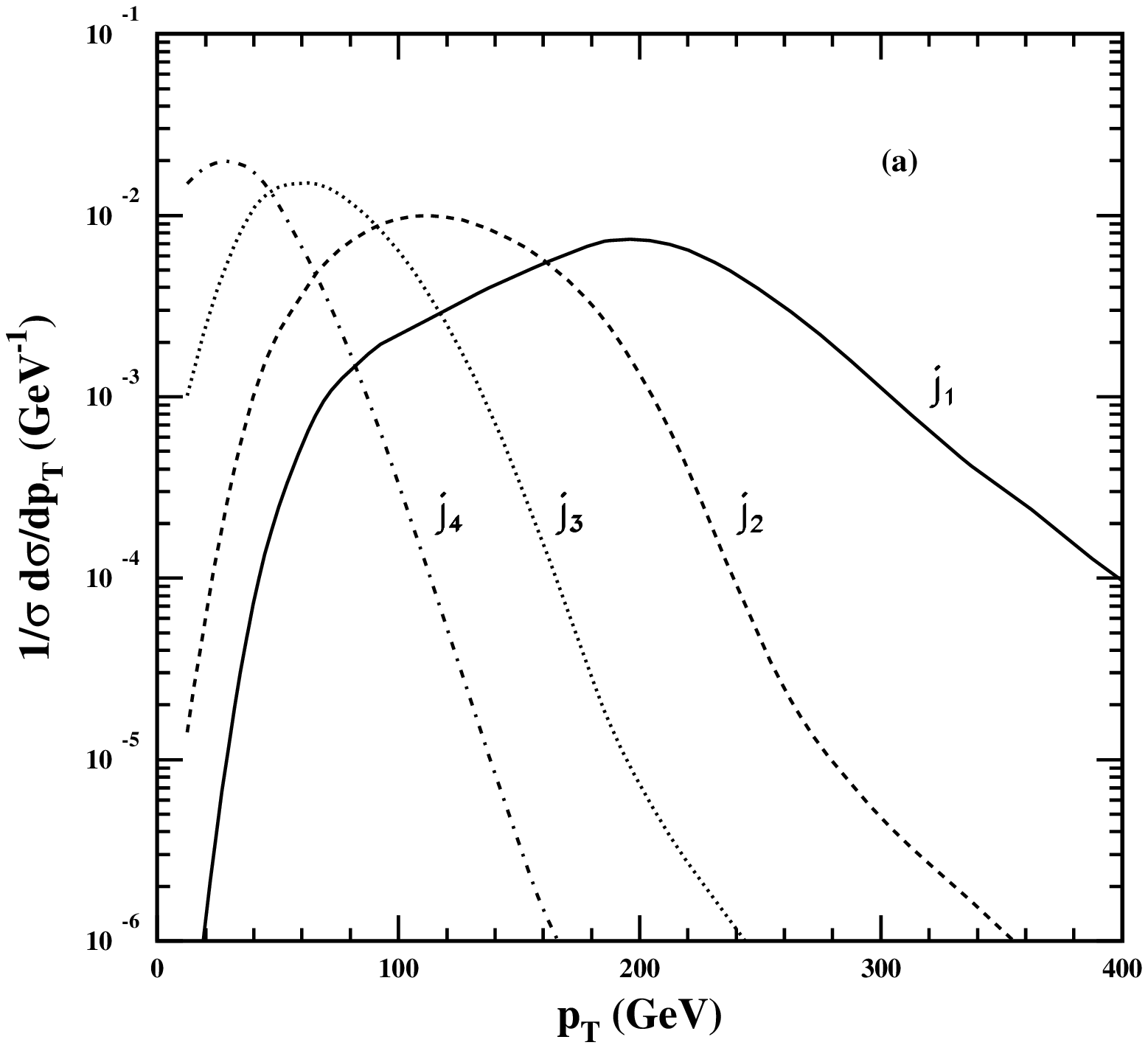} & \includegraphics[clip,width=6cm,height=6cm,keepaspectratio]{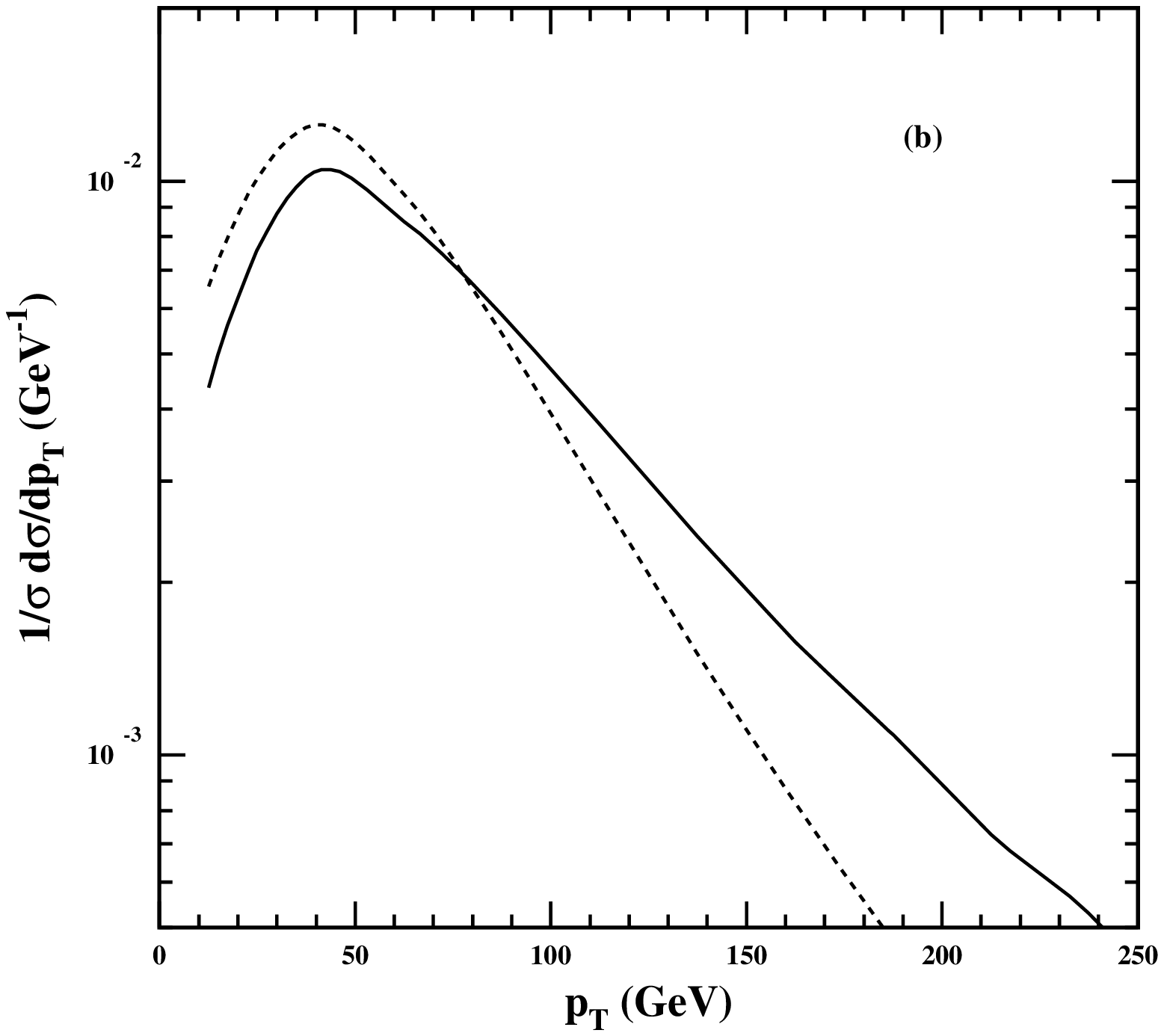} & \includegraphics[clip,width=6cm,height=6cm,keepaspectratio]{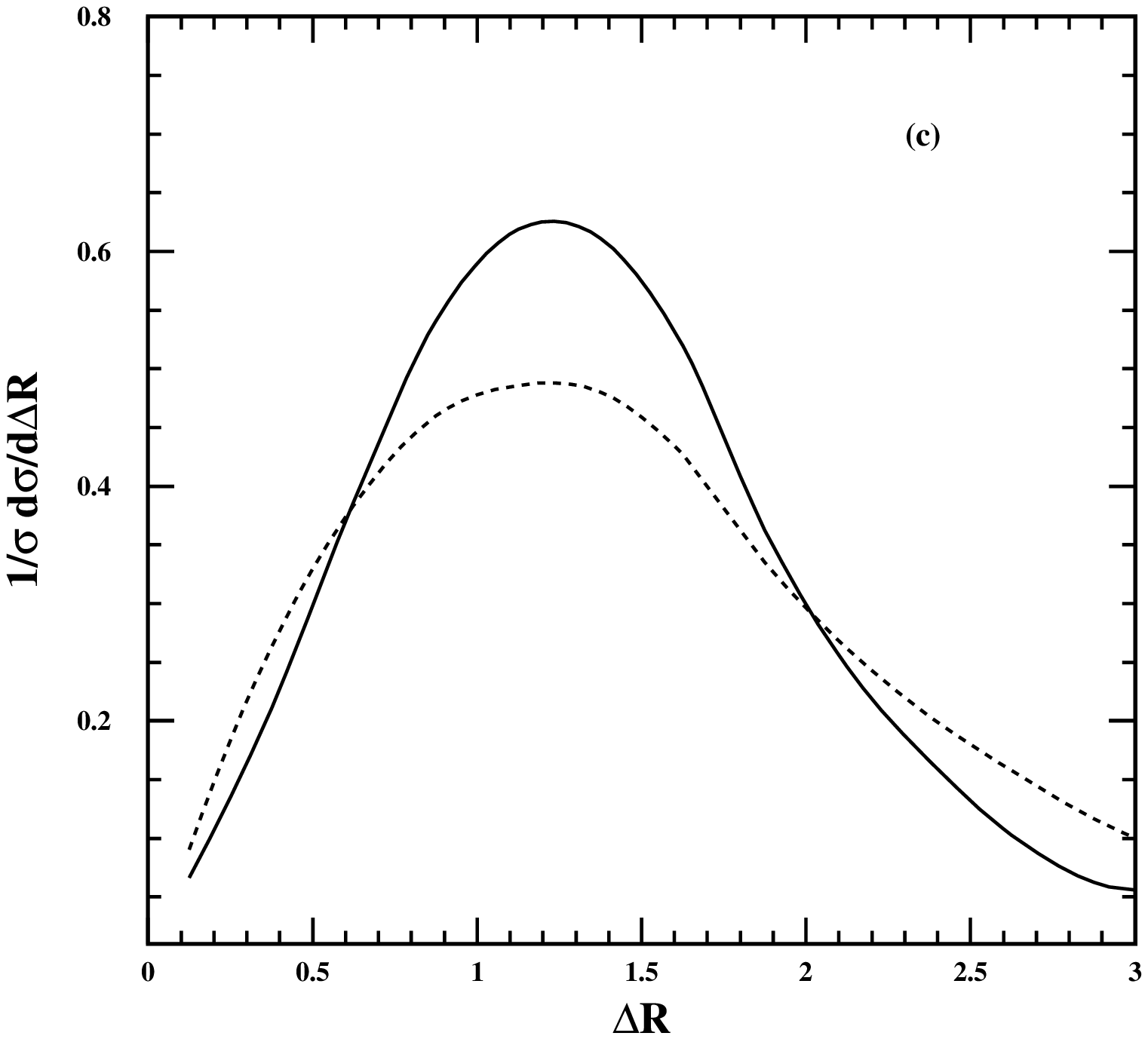} \tabularnewline
\end{tabular}
\caption{(a) The transverse momentum distributions of the jets ($j_1,j_2,j_3,j_4$) with $p_{Tj_1}>p_{Tj_2}>p_{Tj_3}>p_{Tj_4}$ for $m_{H^{\pm}}=500$ GeV at $\sqrt{s}=14$ TeV.
(b) The transverse momentum distribution of the charged lepton (solid
line) and the missing transverse energy $\met$ distribution (dashed line). (c) The
minimal angular separation distributions between jets (solid line) and that between jets and the charged lepton (dashed
line).}
\label{fig:pt}
\par\end{centering}
\centering{} 
\end{figure}
\begin{figure}
\begin{centering}
\includegraphics[clip,width=6cm,height=6cm,keepaspectratio]{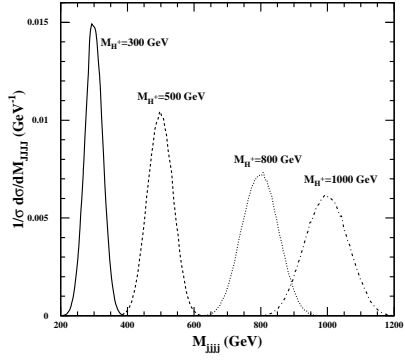}
\caption{The $b\bar b jj$ invariant mass distributions.}
\label{fig:mjjbb}
\par\end{centering}
\end{figure}

To be more realistic, the simulation at the detector is performed
by smearing the leptons and jets energies according to the assumption
of the Gaussian resolution parametrization 
\begin{equation}
\frac{\delta(E)}{E}=\frac{a}{\sqrt{E}}\oplus b,
\end{equation}
where $\delta(E)/E$ is the energy resolution, $a$ is a sampling
term, $b$ is a constant term, and $\oplus$ denotes a sum in quadrature.
We take $a=5\%$, $b=0.55\%$ for leptons and $a=100\%$, $b=5\%$
for jets respectively\cite{Aad:2009wy}.

The transverse momentum distributions for the four jets are shown
in Fig.~\ref{fig:pt} (a). In Fig.~\ref{fig:pt} (b), the transverse
momentum distribution for the charged lepton and the missing transverse energy
($\met$) distribution are displayed. In order to identify the isolated
jet (lepton), we define the angular separation between particle $i$
and particle $j$ as 
\begin{equation}
\Delta R_{ij}=\sqrt{\Delta\phi_{ij}^{2}+\Delta\eta_{ij}^{2}},
\end{equation}
 where $\Delta\phi_{ij}=\phi_{i}-\phi_{j}$ and $\Delta\eta_{ij}=\eta_{i}-\eta_{j}$.
$\phi_{i}$ ($\eta_{i}$) denotes the azimuthal angle (rapidity) of
the related jet (or lepton). The corresponding distributions for $\Delta R=min(\Delta R_{ij})$ are shown
in Fig.~\ref{fig:pt} (c). 

The momentum of the neutrino can be
reconstructed from the $W$ mass, missing transverse momentum and
neutrino mass, i.e., the neutrino momentum is obtained by solving
the following equations
 \begin{eqnarray}
{\bf p}_{\nu T} & = & -({\bf p}_{lT}+\sum{\bf p}_{jT}),\nonumber \\
m_{W}^{2} & = & (p_{\nu}+p_{l})^{2},\; p_{\nu}^{2}=0,
\end{eqnarray}
 where ${\bf p}_{iT}$ is the transverse momentum of the corresponding particle
and $p_{\nu}$ ($p_{l}$) is the four-momentum of neutrino (charged lepton). We veto the event if no solution can be found.

In our analysis, all the hadronic jets are from
the charged Higgs boson decay. As a result, the invariant mass of
these jets can be used to reconstruct the charged Higgs boson
mass. The distributions $d\sigma/dM_{jjjj}$ for various charged
Higgs mass are shown in Fig.~\ref{fig:mjjbb}.

Based on the above discussion, we employ the
basic cuts (referred as cut I)
 \begin{eqnarray}
 & p_{lT}>20~GeV,~p_{jT}>30~GeV,~\met>20~GeV,\nonumber \\
 & |\eta_l|<2.5,~|\eta_j|<3.0,~\Delta R_{jj(lj)}>0.4.
\end{eqnarray}

For the processes Eq.~(\ref{pro:decay}) with final state $ l+\met+b\bar{b}jj$,
the dominant SM backgrounds are $t\bar{t}$, $t\bar{t}W$, $t\bar{t}Z$, $WZjj$, $WWjj$ and $Wjjjj$, 
which are generated with the MadGraph\cite{madgraph}
and Alpgen\cite{alpgen}. In the $W^{\pm}H^{\mp}$ production
processes, four jets are from the charged
Higgs decay, and three of them are from top quark decay.
Therefore to purify the signal, one can require the invariant mass
of final jets to be around the charged Higgs mass, and one top quark
is reconstructed by three jets. Since $t\bar{t}$
is one of the predominant backgrounds, one can veto $t\bar{t}$
events if the second top quark can be reconstructed. Such kinds of
invariant mass cut (referred as cut II) are
\begin{eqnarray}
|M_{jjjj}-m_{H^{\pm}}|\leq30~GeV,~~|M_{jjj}-m_{t}|\leq20~GeV~and~~|M_{jl\nu}-m_{t}|\geq20~GeV.
\end{eqnarray}
 In order to further
purify the signal, we apply a cut on the invariant mass for all of
the visible particles
\begin{eqnarray}
M_{jjjjl}=\sqrt{(\sum p_{j}+p_{l})^{2}}>1.5m_{H^{\pm}}
\end{eqnarray}
together with one b-tagging (referred as cut III). 

The cross sections for 
signal after each cut are listed in table~\ref{tab:cs-signal_14tev}. 
We find that after all cuts, there is about 1 $fb$ left for the signal process around $m_{H^{\pm}}=500$ GeV, and the backgrounds are suppressed significantly. In table~\ref{tab:cs-bgs}, we list the event numbers for the signal and background processes survived after all cuts with the integral luminosity of $300~ fb^{-1}$  at $\sqrt{s}=14$ TeV. The significance for the signal to background can reach above three sigma for $m_{H^{\pm}}\ge 400$ GeV. As an example, we choose $m_{H^{\pm}}=$ 
500 GeV for the investigation in section~\ref{sec:wprime}.

\begin{table}[h]
\begin{centering}
\begin{tabular}{|c|c|c|c|c|}
\hline 
$m_{H^{\pm}}$(GeV)  & No cuts & Cut I  & Cut I+II & Cut I+II+III \tabularnewline
\hline 
300 & 105 & 6.37 & 3.33 & 1.43\tabularnewline
\hline 
400 & 49.7 & 4.41  & 2.05  & 1.11 \tabularnewline
\hline 
500 & 25.7  & 2.83  & 1.20 & 0.72 \tabularnewline
\hline 
600 & 14.1 & 1.81 & 0.70 & 0.45\tabularnewline
\hline 
800 & 4.90 & 0.77 & 0.26 & 0.18\tabularnewline
\hline 
1000 & 1.93 & 0.35 & 0.10 & 0.07\tabularnewline
\hline
\end{tabular}\caption{The cross section of the signal process at  $\sqrt{s}=14$
TeV in unit of $fb$.}
\label{tab:cs-signal_14tev} 
\par\end{centering}

\centering{}

\end{table}

\begin{table}[h!]
\begin{centering}
\begin{tabular}{|c|c|c|c|c|c|c|}
\hline
$m_{H^{\pm}}(GeV)$ &300&400&500&600&800&1000\\ \hline
$W^{\pm}H^{\mp}$& 429&333 &216&135&54&21 \\ \hline
$ t\bar t$&14640 &3360 &696&231&$<$1&$<$1 \\ \hline
$ t\bar t W$& 24&9&3&2&$<$1& $<$1\\ \hline
$ t\bar t Z$& 30&6&$<$1&$<$1&$<$1&$<$1\\ \hline
$WZjj$& 69&$<$1&$<$1&$<$1&$<$1&$<$1 \\ \hline
$WWjj$& 120&117&$<$1&$<$1&$<$1&$<$1\\ \hline
$Wjjjj$&20940  &5970&$<$1&$<$1&$<$1&$<$1\\ \hline
$S/B$& 0.01&0.03&0.31&0.57&$>$9&$>$3.5 \\ \hline
$S/\sqrt{B}$ & 2.27&3.42&8.15&8.77&$>$22&$>$8.57 \\ \hline
\end{tabular}
\caption{The event numbers after all cuts for the signal and backgrounds with a integral luminosity of $300~fb^{-1}$ at $\sqrt{s}=14$ TeV. The signal$-$background ratio and the significance are given.}
\label{tab:cs-bgs} 
\par\end{centering}

\centering{}
\end{table}

\section{Identify $H^{\pm}$  from charged vector bosons}\label{sec:wprime}

The reconstruction of the resonance from $b\bar{t}~(t\bar{b})$
final states can straightforward lead to the conclusion that a new
charged boson is detected. However, many theories beyond SM also predict
the existence of new heavy charged vector bosons (e.g. $W^{\prime\pm}$)
which can decay to $b\bar{t}~(\bar{tb})$. It can not be ignored
to identify the scalar from the vector bosons with the identical
final state. We study the following processes
\begin{eqnarray}
&pp&\to W^{-}W'^{+}\to W^{-}t\bar{b}\to l^{-}\nu b\bar{b}jj, \nonumber \\
&pp&\to W^{+}W'^{-}\to W^{+}\bar{t}b\to l^{+}\nu b\bar{b}jj.
\label{pro:decay}
\end{eqnarray}
The corresponding transverse momentum distributions and $b\bar b jj$ invariant mass are similar to Fig.~\ref{fig:pt} and Fig.~\ref{fig:mjjbb}.
Hence a proper observable has to be found to represent the spin of the new particle, which is another main aim of
this paper.
\begin{figure}
\begin{centering}
\begin{tabular}{cc}
\includegraphics[clip,width=6cm,height=6cm,keepaspectratio]{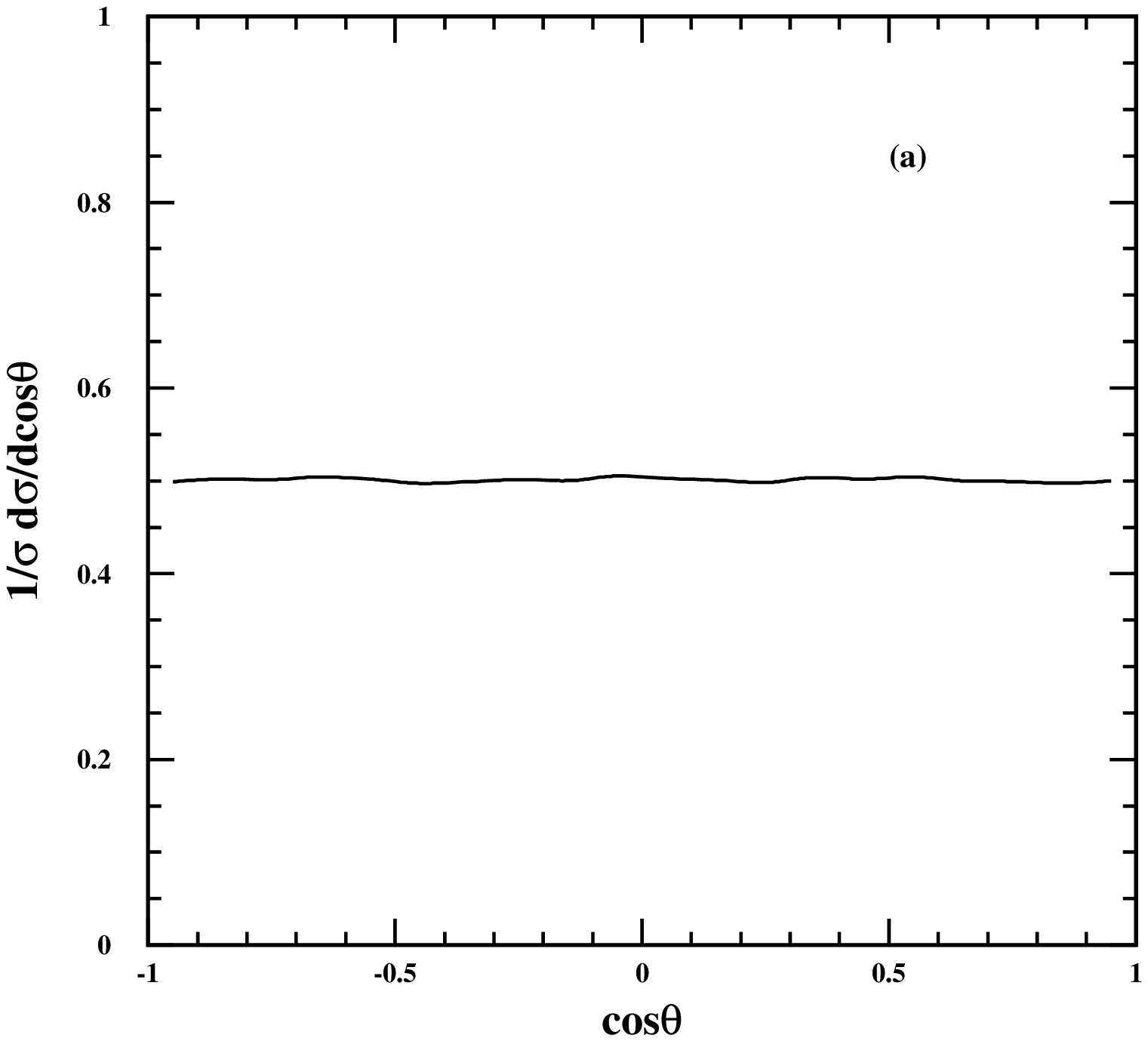}  & 
\includegraphics[clip,width=6cm,height=6cm,keepaspectratio]{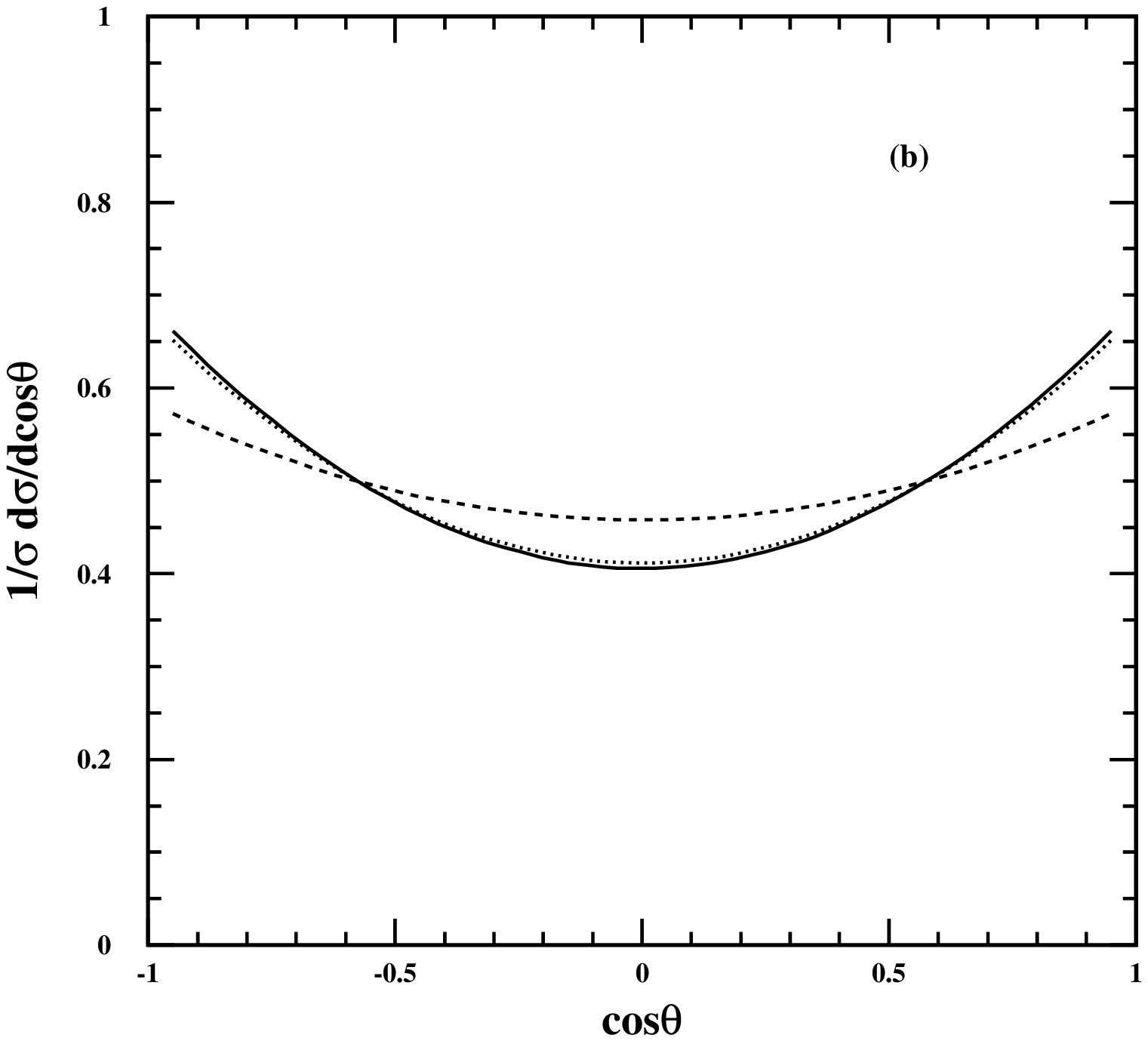}\tabularnewline
\end{tabular}
\caption{The angular distribution of bottom quark in the (a) $W^{\pm}H^{\mp}$
process and (b) $W^{\pm}W'^{\mp}$ process at LHC for $m_{(H^{\pm},W^{\prime\pm})}=500$
GeV. Solid, dashed and dotted lines respectively stand for the distributions
through $W_{L}^{\prime\pm}$, $W_{R}^{\prime\pm}$ and $W_{V}^{\prime\pm}$. The symbols $W_{L}^{\prime \pm}$ ($W_{R}^{\prime \pm}$) and $W_{V}^{\prime \pm}$  respectively denote the vector bosons participated in left (right) handed and pure vector type interactions.}
\label{fig:theta} 
\par\end{centering}
\begin{centering}
\par\end{centering}
\centering{} 
\end{figure}
\begin{figure}
\begin{centering}
\begin{tabular}{cc}
\includegraphics[clip,width=6cm,height=6cm,keepaspectratio]{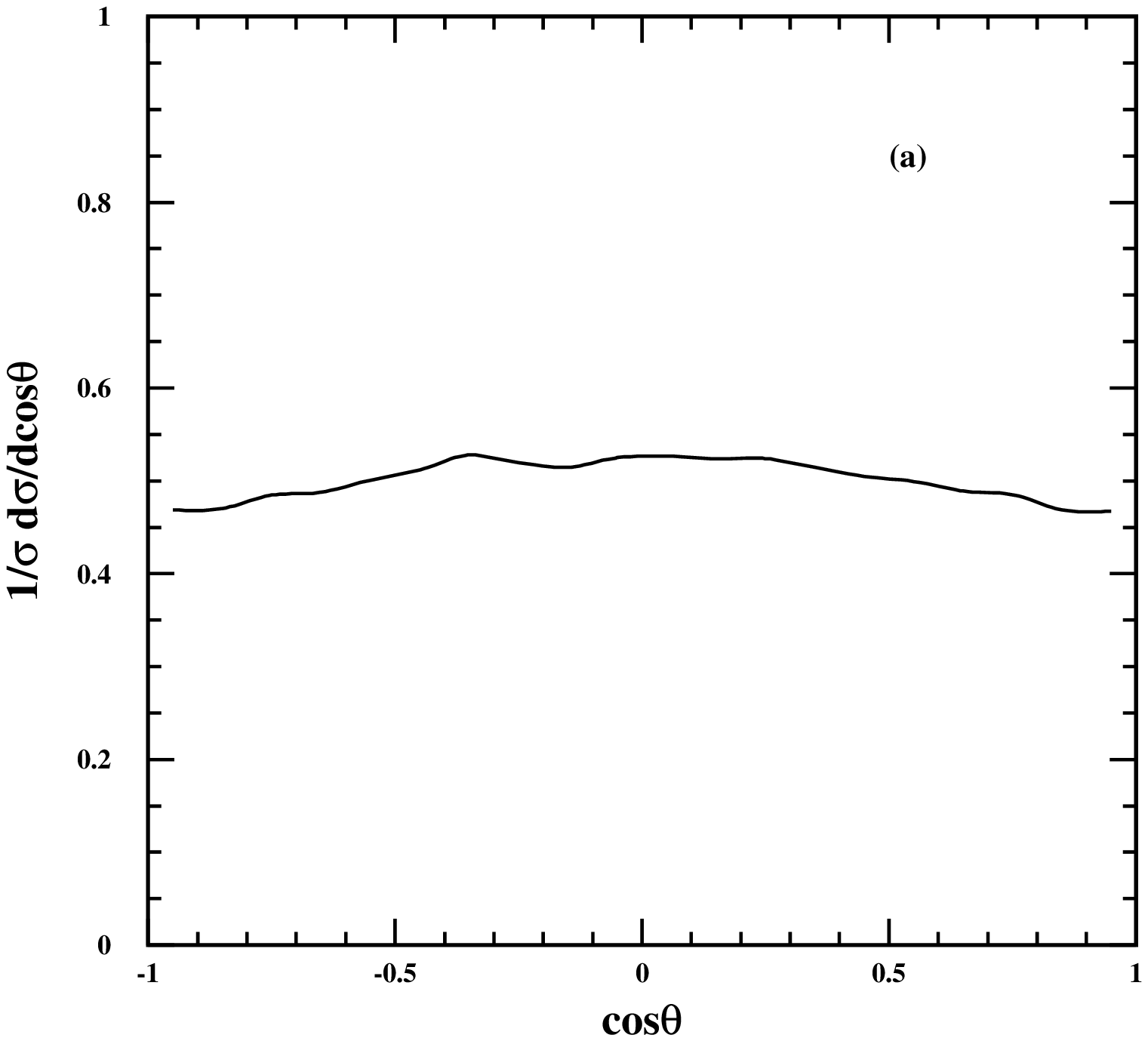}  & 
\includegraphics[clip,width=6cm,height=6cm,keepaspectratio]{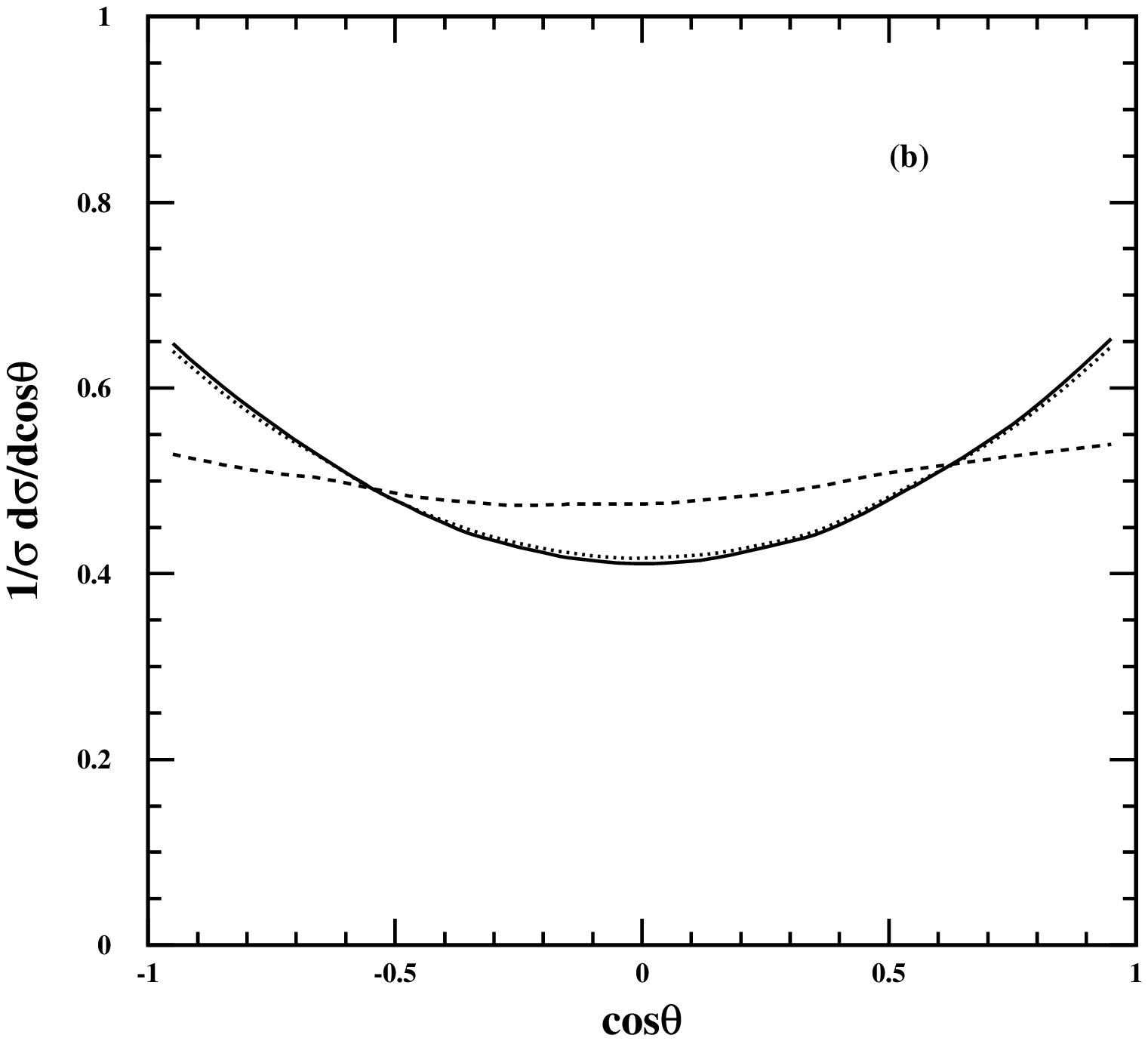}\tabularnewline
\end{tabular}
\caption{The angular distribution of b-jet in the (a) $W^{\pm}H^{\mp}$
process and (b) $W^{\pm}W'^{\mp}$ process at LHC after the detector
simulation and through all cuts, Solid, dashed and dotted lines respectively
stand for the distributions through $W_{L}^{\prime\pm}$, $W_{R}^{\prime\pm}$
and $W_{V}^{\prime\pm}$. The symbols are similar to Fig.~\ref{fig:theta}.}
\label{fig:thetacut} 
\par\end{centering}
\begin{centering}
\par\end{centering}
\centering{} 
\end{figure}
Taking into account that the scalar particle is different from the
vector ones in the angular distribution of their decay products,
we can define the angle $\theta$ as follows 
\begin{equation}
\cos\theta=\frac{{\bf p}_{b}^{*}\cdot{\bf p}}{|{\bf p}_{b}^{*}||{\bf p}|},
\end{equation}
where ${\bf p}$ is the 3-momentum of one of the initial proton in
the laboratory frame, and ${\bf p}_{b}^{*}$ is the 3-momentum of
the b-jet which is not decay from top (anti-)quark in the $H^{\pm}$ (or
$W'^{\pm}$) rest frame. The distributions $\frac{d\sigma}{\sigma d\cos\theta}$
related to $H^{\pm}$ and $W'^{\pm}$ without any cuts are shown in Fig.~\ref{fig:theta}.
The bottom (anti-)quark in $H^{\pm}$ rest frame is isotropic, and
has no correlation with the proton moving direction. However, if the
new charged particle is a vector ($W^{\prime\pm}$), the ${\bf p_{b}^{*}}$
is not isotropic any more. The distribution is sensitive to the chiral couplings. 
Such chiral couplings of $W^{\prime \pm}$ have also been studied in other processes\cite{Bao:2011nh,Gopalakrishna:2010xm}. 
Fig. \ref{fig:thetacut} shows the corresponding
b-jet angular distributions after the smearing and the kinematic cuts
in section~\ref{signal}. One can find that the curve for $H^{\pm}$
is slightly distorted by the kinematic cuts, which does not change the fact that the distributions corresponding to the scalar and the vector bosons are characteristically different. 
To characterize such difference, we employ the function 
\begin{equation}
f(\cos\theta)=\frac{1}{2+\frac{2}{3}A}(1+A\cos^{2}\theta)
\label{fitfunction}
\end{equation}
to fit the curves in Fig.~\ref{fig:theta} and \ref{fig:thetacut}.
The values of $A$ are listed in table \ref{tab:fit}. Obviously, $A$ is different for scalar and vector bosons before or after all cuts. Therefore, the investigation of the angular distribution and the characteristic quantity $A$ is helpful to discriminate the charged scalar from the
vector bosons.
\begin{table}[h!]
\begin{centering}
\begin{tabular}{|c|c|c|c|c|}
\hline 
 & $H^{\pm}$ & $W_{L}^{\prime\pm}$ & $W_{R}^{\prime\pm}$ & $W_{V}^{\prime\pm}$\tabularnewline
\hline 
No cuts & 0 & 0.69 & 0.28 & 0.65\tabularnewline
\hline 
Cut I+II+III & -0.13 & 0.66 & 0.14 & 0.60\tabularnewline
\hline 
\end{tabular}\caption{The values of $A$ from fitting the curves in Fig.~\ref{fig:theta} and \ref{fig:thetacut}
with the function Eq.~(\ref{fitfunction}).}
\label{tab:fit}
\par\end{centering}

\centering{}
\end{table}

\section{Summary}

We investigate the possibility of detecting charged Higgs
production in association with $W$ boson via $pp\to W^{\pm}H^{\mp}\to l+\met+b\bar{b}jj$ at LHC.
 We apply the resonance reconstruction (cut II) and  resonance mass dependence (cut III) to suppress
the QCD backgrounds. It is found that with $300~fb^{-1}$ integral
luminosity at $\sqrt{s}=14$ TeV, the signal can be distinguished from the backgrounds for the charged Higgs
mass around 400 GeV or larger.
If a new resonance particle is observed, 
one of the key question is to identify its spin.
We study the angular distributions for the charged Higgs and the representative
vector bosons ($W^{\prime \pm}$) with various chiral couplings. From the angular distributions
and the characteristic quantity $A$ the scalar can be distinguished from the vector bosons. 
The above analysis can be applied to discriminate the scalar from the vector bosons once the
new resonance particle produced in association with a vector gauge boson (e.g. $W$ or $Z$) is
discovered at LHC.

\begin{acknowledgments}

This work was supported in part by the National Science Foundation
of China (NSFC), Natural Science Foundation of Shandong Province (JQ200902, JQ201101). The authors thank all of
the members in Theoretical Particle Physics Group of Shandong University for their helpful discussions.
\end{acknowledgments} 


\end{document}